\documentclass[final]{IEEEtran}

\ifCLASSINFOpdf
   \usepackage[pdftex]{graphicx}
\else
   \usepackage[dvips]{graphicx}
\fi

\usepackage[nolist]{acronym}
\usepackage[cmex10]{amsmath}
\usepackage{amssymb}
\usepackage{gensymb}

\usepackage{url}

\makeatletter
\let\old@ps@headings\ps@headings
\let\old@ps@IEEEtitlepagestyle\ps@IEEEtitlepagestyle
\def\psccfooter#1{%
    \def\ps@headings{%
        \old@ps@headings%
        \def\@oddfoot{\strut\hfill#1\hfill\strut}%
        \def\@evenfoot{\strut\hfill#1\hfill\strut}%
    }%
    \def\ps@IEEEtitlepagestyle{%
        \old@ps@IEEEtitlepagestyle%
        \def\@oddfoot{\strut\hfill#1\hfill\strut}%
        \def\@evenfoot{\strut\hfill#1\hfill\strut}%
    }%
    \ps@headings%
}
\makeatother

\begin{document}

\title{Data-Driven Demand-Side Flexibility Quantification: Prediction and Approximation of Flexibility Envelopes\\
\thanks{This work has been submitted to the IEEE for possible publication. Copyright may be transferred without notice, after which this version may no longer be accessible.}}

\author{
\IEEEauthorblockN{
Nami Hekmat\IEEEauthorrefmark{1},
Hanmin Cai\IEEEauthorrefmark{2},
Thierry Zufferey\IEEEauthorrefmark{1}, 
Gabriela Hug\IEEEauthorrefmark{1}, and
Philipp Heer\IEEEauthorrefmark{2}
}\\

\IEEEauthorblockA{\IEEEauthorrefmark{1} Power Systems Laboratory, ETH Zurich, Switzerland, \{nhekmat, thierryz, hug\}@ethz.ch}\\

\IEEEauthorblockA{\IEEEauthorrefmark{2} Urban Energy Systems Laboratory, Empa, Switzerland, \{hanmin.cai, philipp.heer\}@empa.ch}
}

\begin{acronym}
\acro{ANN}{artificial neural network}
\acro{CDF}{cumulative distribution function}
\acro{DSO}{distribution system operator}
\acro{EV}{electric vehicle}
\acro{GMM}{Gaussian mixture model}
\acro{HP}{heat pump}
\acro{KNN}{k-nearest neighbors}
\acro{MAE}{mean absolute error}
\acro{ML}{machine learning}
\acro{MPC}{model predictive control}
\acro{ND}{normal distribution}
\acro{NRMSE}{normalized root-mean-square error}
\acro{PDF}{probability distribution function}
\acro{PV}{photovoltaics}
\acro{RMSE}{root-mean-square error}
\acro{SH}{space heating}
\acro{SND}{skewed normal distribution}
\acro{SVM}{support vector machine}
\acro{SVR}{support vector regression}
\end{acronym}

\maketitle

\begin{abstract}
Real-time quantification of residential building energy flexibility is needed to enable a cost-efficient operation of active distribution grids.
A promising means is to use the so-called \textit{flexibility envelope} concept to represent the time-dependent and inter-temporally coupled flexibility potential. However, existing optimization-based quantification entails high computational burdens limiting flexibility utilization in real-time applications, and a more computationally efficient approach is desired. Also, the communication of a flexibility envelope to system operators in its original form is data-intensive. To address these issues, this paper first trains several machine learning models based on historical quantification results for online use. Subsequently, probability distribution functions are proposed to approximate the flexibility envelopes with significantly fewer parameters, which can be communicated to system operators instead of the original flexibility envelope. The results show that the most promising prediction and approximation approaches allow for a minimum reduction of the computational burden by a factor of 9 and of the communication load by a factor of 6.6, respectively.
\end{abstract}


\acresetall

\section{Introduction}\label{sec:introduction}
The penetration of distributed energy resources such as \ac{PV}, \acp{HP}, and \acp{EV} in low-voltage distribution networks is growing, thereby increasing the stress in the system and potentially resulting in voltage band violations and congestion. 
To address these issues cost-efficiently, leveraging building energy flexibility may be favored over reinforcing the existing infrastructure \cite{spiliotis2016demand}. In this work, flexibility refers to the capability of modifying energy usage schedules without violating operational constraints or compromising occupants' comfort. 

To utilize flexibility at the building level, previous studies have proposed predictive energy management systems \cite{cai2021experimental}. Besides optimizing the energy flow, buildings are further suggested to quantify flexibility potentials, which are periodically reported to \acp{DSO} for system-level coordination. 

In this paper, a \textit{flexibility envelope} concept quantifying building energetic flexibility is considered \cite{cai2021experimental}. It adopts a model-based approach and characterizes flexibility as a three-dimensional surface spanned by lead time, feasible power levels, and the corresponding maximum sustained duration. In Fig.~\ref{fig:envelope}, an example is provided. 
However, rolling updates of a flexibility envelope require solving similar optimization problems with different initial and boundary conditions repeatedly. Although the original optimization problem can be solved reasonably fast with commercial iterative solvers, it entails high computational burdens and limits the response speed. 
Also, it consists of a notable number of data points, and periodical reporting of envelopes from multiple buildings represents a significant communication load.

These issues can be remedied by applying \ac{ML} techniques, which have proved to be successful function approximators. For example, \ac{ML} has found applications in efficient power system security assessments~\cite{thams2019efficient} and approximating optimal control policies~\cite{yang2021experiment}. Both greatly improve computational efficiency, as only function evaluations are needed once \ac{ML} models are trained.
Similarly, we apply popular \ac{ML} techniques to mitigate computational burdens. 
To reduce the communication load, we approximate flexibility envelopes with \acp{PDF}. Only a few hyperparameters need to be transmitted, and the receiver can reasonably recover the original data.
We consider different variants of Gaussian (normal) distributions. These are often applied to approximate the distribution of energy-related data such as residential load profiles~\cite{zufferey2018generating}.

Therefore, the contributions of this paper are twofold. We first train \ac{ML}-based approximators based on historical quantification results to replace the existing repeated optimization at the online stage. The data-intensive representation of flexibility is further approximated using \acp{PDF}, leading to fewer parameters to be communicated with system operators. 

\begin{figure}
    \centering
    \includegraphics[width=\columnwidth]{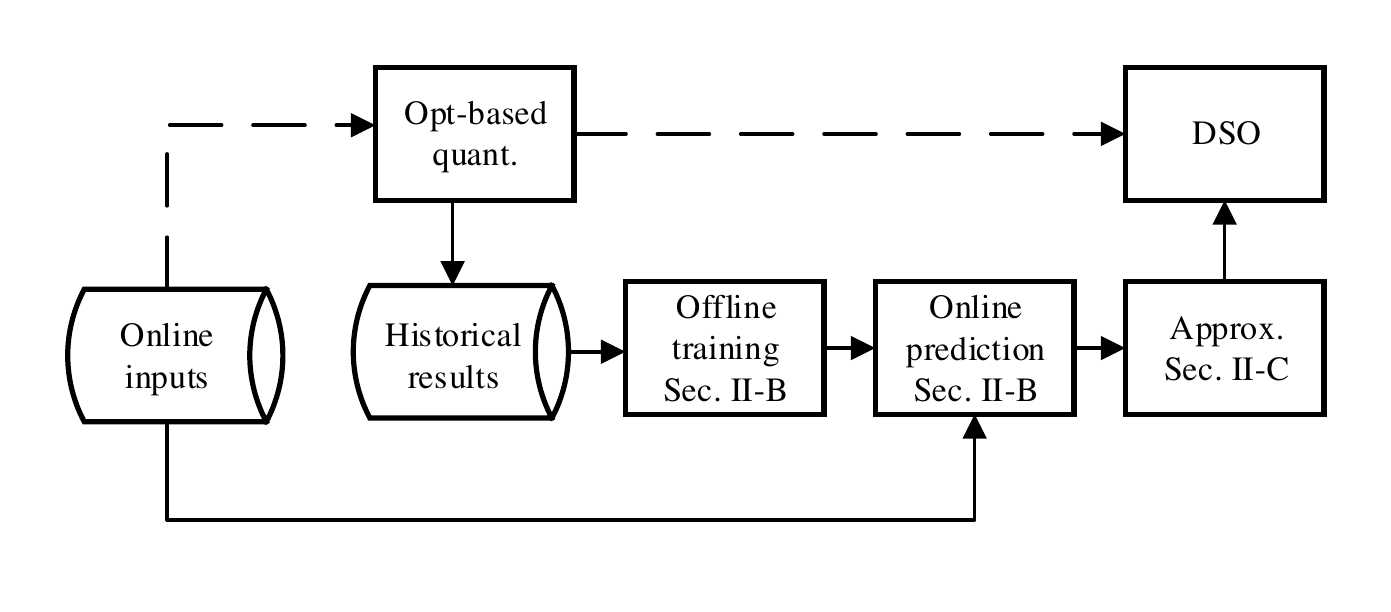}
    \caption{Comparison between the existing optimization-based workflow (in dashed lines) and the proposed workflow (in solid lines).}
    \label{fig:flow_chart}
\end{figure}

The remainder of the paper is organized as follows. In Section \ref{sec:methodology}, we introduce the existing quantification approach, and the \ac{ML} algorithms used for the flexibility envelope prediction and the approximation based on \ac{PDF} (see Fig.~\ref{fig:flow_chart} for a visual summary). In Section \ref{sec:case_study}, we present the data set collected with the existing approach. Results of both the prediction and the approximation of the flexibility envelope are presented in Section \ref{sec:results}. Finally, concluding remarks, including future work, are summarized in Section \ref{sec:conclusion}.

\section{Methodology}\label{sec:methodology}

This section first reviews the existing quantification approach described in \cite{cai2021experimental} and then presents the methodology behind the prediction and approximation of the flexibility envelopes. In both cases, different models are presented and will be compared in Section~\ref{sec:results}.

\subsection{Existing Flexibility Envelope Quantification}\label{sec:original_flex_quant}

\begin{figure}
    \centering
    \includegraphics[width=\columnwidth]{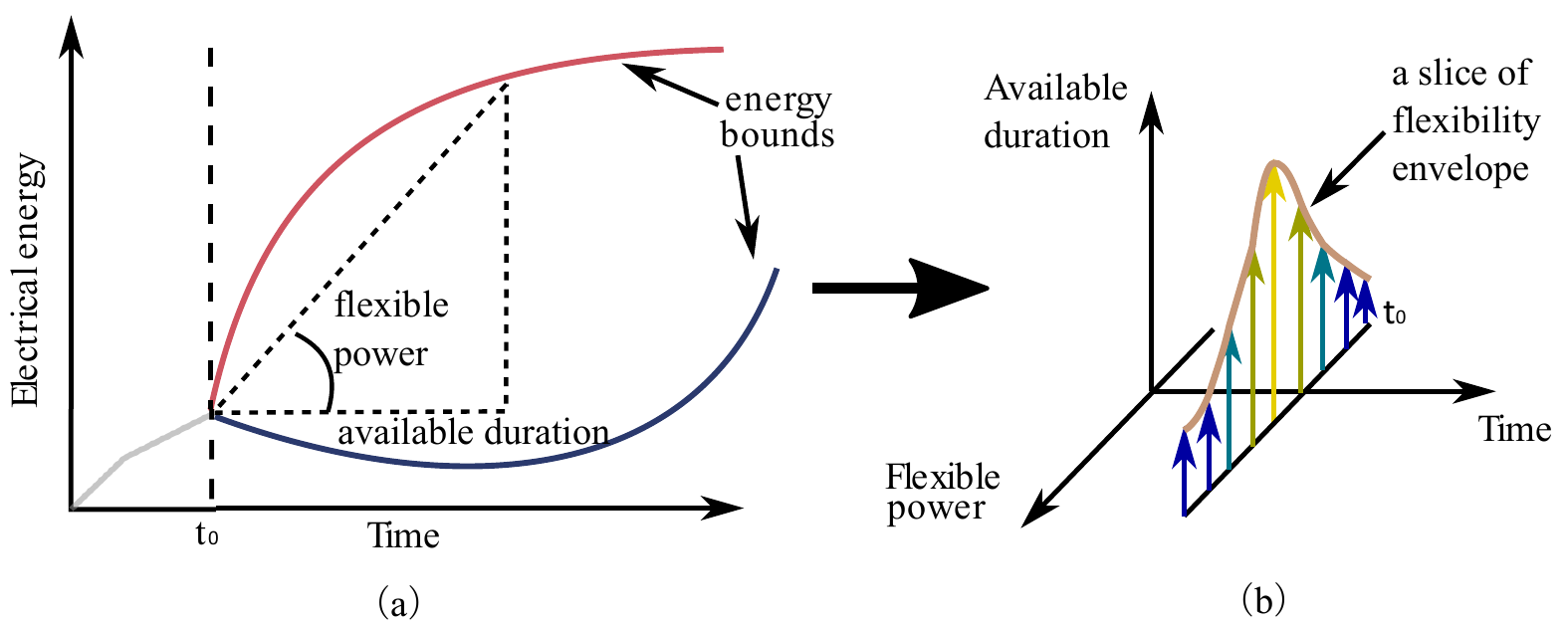}
    \caption{Calculation of maximal sustainable duration by (a) first identifying energy bounds and then (b) deriving sustainable duration of chosen power levels.}
    \label{fig:envelope_calculation}
\end{figure}

Initially, the upper and lower energy bounds are identified by energizing the building to its extremes, which consists of solving two optimization problems as follows \cite{cai2021experimental}:
\begin{subequations}\label{eq:opt_based_quant}
\begin{alignat}{1}
    J_{\mathrm{upper}} &= \min_{\mathbf{u},\delta} \; - \mathbf{w}^\intercal \mathbf{u} + \lVert \mathbf{\delta} \rVert^2_{\mathbf{w}_\delta},\\
    J_{\mathrm{lower}} &= \min_{\mathbf{u},\delta} \; \mathbf{w}^\intercal \mathbf{u} + \lVert \mathbf{\delta} \rVert^2_{\mathbf{w}_\delta},
\end{alignat}
\end{subequations}
where $\mathbf{u}$ is a vector of active power of all components, $\mathbf{w}$ is a vector of weights, $\mathbf{\delta}$ is a vector of slack variables introduced to ensure feasibility and $\mathbf{w}_\delta$ is the corresponding weighting matrix. Optimization constraints include the dynamics of each component, power and energy capacities, state limits and mutual exclusiveness of charging and discharging. The respectively obtained $\mathbf{u}$ gives the upper and lower energy bounds. 
The feasible power levels and the corresponding maximum sustained duration can be derived using the procedure shown in Fig.~\ref{fig:envelope_calculation} \cite{cai2021experimental}. When applying this derivation in all the lead times within the horizon, a flexibility envelope spanned by lead time, power levels, and maximum sustained duration is obtained.  An example of flexibility envelope with a 24-hour horizon is shown in Fig.~\ref{fig:envelope}.
In real-time applications, the entire quantification process is periodically repeated to account for the most recent observations.
In the rest of this paper, the prediction and approximation of flexibility envelopes concretely refer to the prediction and approximation of the maximum sustained duration values, which depend on the two other dimensions.

\subsection{Prediction of Flexibility Envelope}

Each flexibility envelope can be predicted by~\ac{ML} models to replace the computationally intensive approach described in (\ref{eq:opt_based_quant}). 
In this work, a simple benchmark model is compared to a linear regressor, a~\ac{KNN} algorithm, a~\ac{SVM}, and an AdaBoost regressor from the scikit-learn Python library. They are briefly presented in the following subsections. Hyperparameter tuning is performed using grid search with 5-fold cross-validation on the training set.
Predictions are performed based on the following input features:
\begin{itemize}
    \item calendar data: time of the day (cyclic encoding)
    \item weather forecast from MeteoSwiss: solar irradiance (in~$\text{W/m}^2$) and temperature (in~$\degree \text{C}$)
    \item residential behavior from historical data: presence at home (binary), non-flexible load (in~$\textrm{kW}$), water draw (in~$\textrm{L/min}$)
    \item comfort requirements: room temperature limits (in~$\degree \text{C}$), domestic hot water temperature limits (in~$\degree \text{C}$)
    \item predicted energy state of flexible appliances: home battery energy (in~$\textrm{kWh}$),~\ac{EV} battery energy (in~$\textrm{kWh}$), \ac{PV} energy production (in~$\textrm{kWh}$), room temperature  (in~$\degree \text{C}$), domestic hot water tank temperature  (in~$\degree \text{C}$)
\end{itemize}
Since the input features are part of the information required by the existing approach, no additional data are necessary. Furthermore, the linear regressor, KNN, and SVM algorithms require features scaling, which is done by standardization.

\subsubsection{Linear Regressor}

The linear model outputs a linear combination of the input features:
\begin{equation} \label{eq:lin}
	 \hat{y} = \sum_{i=1}^m \beta_i\cdot x_i
\end{equation}
where $\hat{y}$ is the predicted value, $x_i$ is the corresponding $i^{\textrm{th}}$ input feature, $\beta_i$ is the weight associated with the $i^{\textrm{th}}$ input feature, and $m$ is the total number of features. The weights are tuned using the Lasso regression, which consists of minimizing a least-square loss function with L1 regularization:
\begin{equation}\label{eq:lin_loss}
    J_{\mathrm{lin}} = \min_{\beta} \; \lVert y - X\beta \rVert^2 + \gamma \cdot \lVert \beta \rVert
\end{equation}
where $y$ is a vector of length $n$ containing the target values in the training set, $X$ is an $n \times m$-matrix containing the input features, $\beta$ is a vector of length $m$ containing the weights of the linear regression, and $n$ is the total number of examples in the training set. The regularization parameter $\gamma$ is a hyperparameter whose value is selected from $\{0.2, 0.4, 0.6, 0.8, 1\}$.

\subsubsection{K-Nearest Neighbors}

The~\ac{KNN} is a non-parametric model whose prediction is a linear combination of the target values of the $k$ nearest neighbors in the training set:
\begin{equation} \label{eq:knn}
	 \hat{y} = \sum_{i=1}^k w_i^{\mathrm{knn}}\cdot y^{\mathrm{knn}}_i,
\end{equation}
where $\hat{y}$ is the predicted value, and $y^{\mathrm{knn}}_i$ and $w_i^{\mathrm{knn}}$ are the target value and the weight associated to the nearest neighbor $i$, respectively. In this work, the model comes from the KNeighborsRegressor Python class, where the number of nearest neighbors $k$ is a hyperparameter whose value is selected from $\{20, 40, 60, 80, 100, 120, 140, 160\}$. Standard Euclidean distance is used as the similarity metric.

\subsubsection{Support Vector Machine}

Used as a regressor, the~\ac{SVM} is also a non-parametric approach that consists of two steps.  In the first step, the original input features are mapped into a higher dimensional space induced by a so-called kernel function. In this new space, the computation occurs in a more efficient manner. In the second step, the algorithm determines the hyper-plane with the narrowest possible margin with a so-called $\varepsilon$-insensitive loss through the input data. There is an extensive literature on the SVR algorithm and more information can be found in~\cite{drucker1997support}. In this work, the model comes from the SVR Python class, where $\varepsilon$ is set to $0.1$ and the radial basis function is chosen as kernel function. Furthermore, the regularization parameter $C$ is a hyperparameter whose value is selected from $\{0.1, 0.2, 0.5, 1, 2, 5\}$.

\subsubsection{AdaBoost Regressor}

The AdaBoost approach is an ensemble method (i.e., multiple weak learners are combined to create a strong learner), where the training of the $(n+1)^\textrm{th}$ weak learner depends on the results of the $n^\textrm{th}$ weak learner. Hence, each weak learner focuses on the error of the previous weak learner. The final prediction is a weighted average of the prediction of all weak learners, where the respective weights are calculated based on the error of each weak learner. More information on the AdaBoost regressor is available in~\cite{freund1997decision}. In this work, the model comes from the AdaBoostRegressor Python class, and the weak learners consist of decision tree regressors. The number of weak learners and the maximal depth of decision trees are hyperparameters whose values are selected from $\{20, 50, 100\}$ and $\{3, 5, \textrm{none}\}$, respectively.

\subsubsection{Benchmark Model}

The \ac{ML}-based predictors are also compared to a simple benchmark model which exploits the time-dependency and the daily pattern of flexibility. Concretely, the benchmark model divides the day into intervals of the same length. For each interval, the average sustained duration is computed for each power level based on the data in the training set. These average values are used as predictions for the sustained duration in the test set:
\begin{equation} \label{eq:benchmark}
	 \hat{y}_j = \frac{1}{|\Omega|}\sum_{i\in\Omega} y_i, \quad \forall j\in\Omega,
\end{equation}
where $\hat{y}_j$ and $y_i$ are the predicted and historical values whose characteristics comply with the condition of $\Omega$, i.e. correspond to a given period of the day. The length of the intervals is a hyperparameter selected from $\{0.25\textrm{h}, 0.5\textrm{h}, 1\textrm{h}, 2\textrm{h}, 4\textrm{h}, 8\textrm{h}\}$ and varies among the various power levels.

\subsection{Approximation of Flexibility Envelope}

The approximation of flexibility envelopes via~\acp{PDF} allows for a reduction of the number of parameters to be transmitted to \acp{DSO}, however at the cost of decreased accuracy. In this work, the flexibility envelopes are approximated based on the 2-dimensional (2D)~\ac{ND}, the 2-dimensional~\ac{SND}, and the 3-dimensional~\ac{GMM}. The 2D models fit a \ac{PDF} to a slice of the flexibility envelope, i.e. at a given lead time within the horizon (see Fig.~\ref{fig:envelope_calculation}(b)), whereas the 3D model directly approximates the entire envelope. Here, the horizon refers to the entire time window (i.e., set of all considered lead times) of the flexibility envelope.

\subsubsection{2D Normal Distribution}

A flexibility envelope at a given lead time tends to follow a ~\ac{ND}. In this work, the approximation model consists of a weighted sum of two~\acp{ND}, which shows improved performance over a single~\ac{ND}. Formally, the model is defined as:
\begin{equation}\label{eq:2Dsum}
    f(x)  = \sum_{i=1}^2 c_i \cdot \phi_i(x),
\end{equation}
where $\phi_i(x)$ is a~\ac{ND} evaluated at $x$, and $c_i$ is the corresponding weight. Parameters are optimized by the~\textit{curve.fit} function from the scipy.optimize Python library, which relies on the Levenberg-Marquardt algorithm~\cite{more1978levenberg}.

\subsubsection{2D Skewed Normal Distribution}

Flexibility envelopes appear to be mostly asymmetric. Indeed, buildings import electricity most of the time such that the flexibility potential is generally skewed to the positive side. Hence, the skewed normal distribution is also considered:
\begin{equation}\label{eq:snd}
    \phi^\textrm{skew}(x)  = \frac{2}{\omega} \cdot \phi\left(\frac{x - \xi}{\omega}\right) \cdot \Phi\left(\alpha \cdot \frac{x - \xi}{\omega}\right),
\end{equation}
where $\phi(x)$ is the~\ac{PDF} of the standard~\ac{ND} and $\Phi(x)$ is the~\ac{CDF} of the~\ac{ND}. In addition, $\alpha$, $\xi$, and $\omega$ represent the skewness, the location, and the scale parameters, respectively.

Similar to the previous approximation model, this model considers the superposition of two~\acp{PDF} as defined in~(\ref{eq:2Dsum}), where $\phi_i$ is replaced by $\phi_i^\textrm{skew}$.

\subsubsection{3D Gaussian Mixture Model}

A~\ac{GMM} is the linear combination of a finite number of normal distribution components. Its 3D version can be expressed as follows:
\begin{equation}\label{eq:3D-GMM}
    \textrm{GMM}^{\textrm{3D}}(x)  = \sum_{i=1}^K c_i \cdot \phi^{\textrm{3D}}_i(x),
\end{equation}
where $\phi^{\textrm{3D}}_i(x)$ is the $i\textsuperscript{th}$ 3D Gaussian component, and $c_i$ is the respective weight. In contrast to its 2D version, each 3D Gaussian component is defined by 2 means and a $2\times 2$ covariance matrix. In this work, the covariance matrix is diagonal such that only 2 parameters are necessary. In addition, the number of components $K$ can be set such as $18$ to balance between approximation accuracy and computational cost. Equation~(\ref{eq:3D-GMM}) is an extension of the model presented in~(\ref{eq:2Dsum}). Furthermore, higher accuracy is achieved by duplicating the values at the first and the last lead time prior to model fitting.

\section{Case study}\label{sec:case_study}

The historical data set used is based on an experiment performed at the NEST building, in Switzerland \cite{cai2021experimental}. Considered flexible components include a \ac{PV} installation of $7.3$~kWp, an \ac{EV} of $7$~kW/$50$~kWh, a home battery of $5$~kW/$17.5$~kWh, space heating demand of $4$~kW, and domestic hot water heating demand of $5.7$~kW. Flexibility envelopes were quantified every 15 minutes with the method described in Section~\ref{sec:original_flex_quant}.
We chose an interval of $0.5$~kW when discretizing the feasible power ranging from $-12$~kW to $14$~kW.
Each component was measured every minute and the weather forecast was updated every $12$ hours with an hourly resolution. All the measurements and external forecasts are re-sampled into $15$ minutes interval. The flexibility quantified from this experiment is used as ground-truth for both training and testing. 

\begin{figure}
    \centering
    \includegraphics[width=\columnwidth]{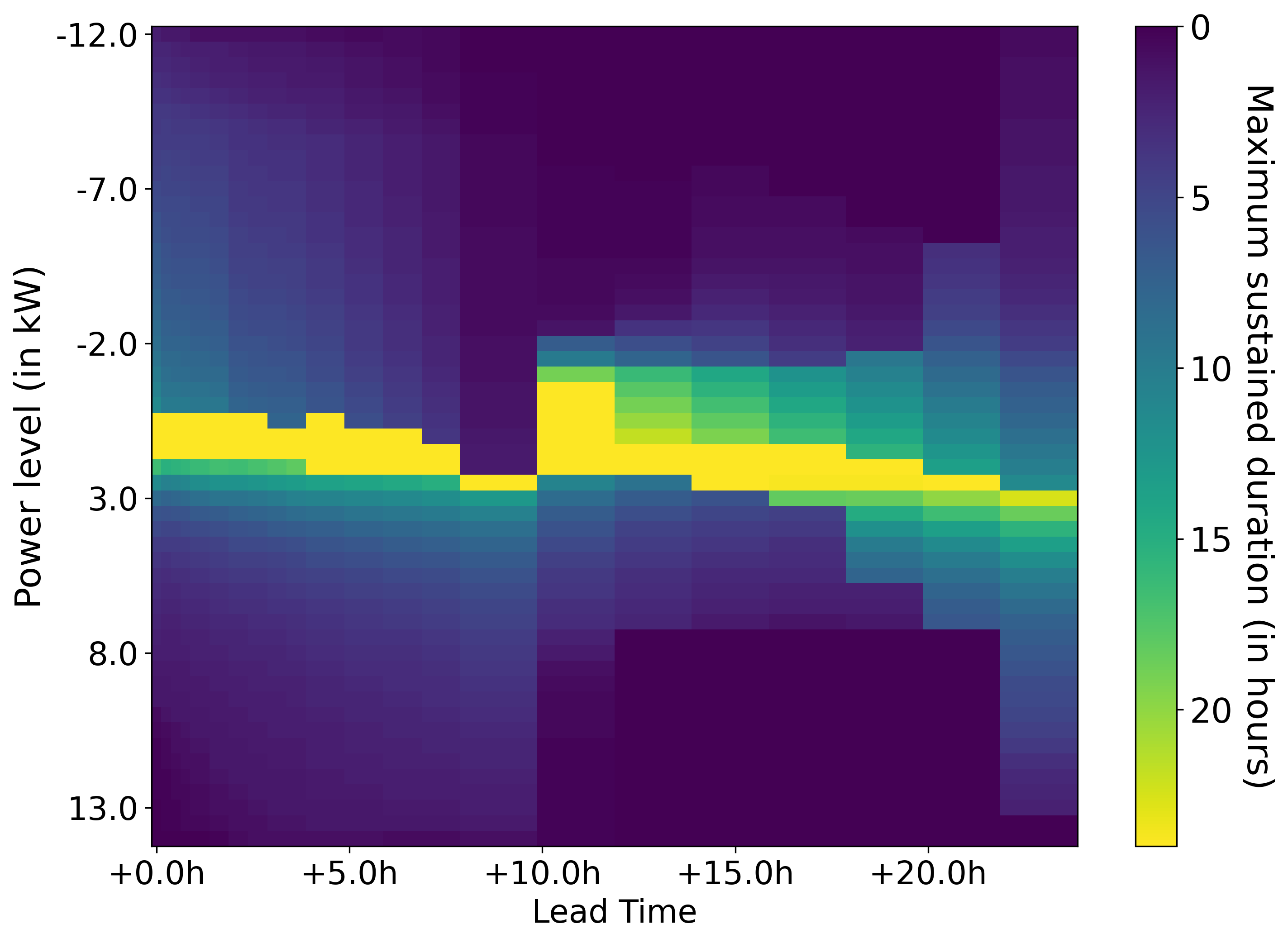}
    \caption{Example of flexibility envelope.}
    \label{fig:envelope}
\end{figure}

Concretely, the case study relies on $547$ flexibility envelopes quantified during working days~\footnote{Large disparities have been observed between working days and weekend. Due to a lack of data for proper training, flexibility envelopes from the weekend are not presented in the case study.}. As illustrated by Fig.~\ref{fig:envelope}, each envelope consists of $53$ power levels. It should be noted that the field experiment considered a set of heterogeneous sampling time intervals over the 24-hour horizon to reduce the computational complexity. Larger sampling time intervals are used for the time instants further in the optimization horizon, resulting in $22$ distinct time steps over the horizon. Hence, one flexibility envelope contains $53\cdot22=1166$ parameters.

\section{Results}\label{sec:results}

This section reports the performance of the prediction and approximation models detailed in Section~\ref{sec:methodology}. Both approaches (i.e., envelope prediction and approximation) are considered independently before assessing the best combination. Finally, computational aspects are discussed.

\subsection{Prediction of Flexibility Envelope}

\begin{table}
\caption{Performance summary of the prediction models}
\label{tab:prediction_summary}
\centering
 \begin{tabular}{|c|c|c|c|}
 \hline
 \begin{tabular}{@{}c@{}}Predict.\\model\end{tabular} &
 \begin{tabular}{@{}c@{}}MAE\\(min)\end{tabular} &
 $\textrm{R}^2$ score &
 \begin{tabular}{@{}c@{}}Comput.\\time (sec)\end{tabular}\\ [0.5ex] 
 \hline
 Original & - & - & $20.75\pm 16.05$  \\
 \hline
 Benchmark & 70 & 0.67 & \textbf{0.15} \\
 Linear & 84 & 0.49 & 0.19\\
 KNN & 42 & 0.78 & 0.59\\
 SVM & 42 & 0.81 & 1.48\\
 AdaBoost & \textbf{25} & \textbf{0.89} & 0.52\\
 \hline
\end{tabular}
\end{table}

The benchmark model and the four \ac{ML}-based models have been tuned on the training set and applied to the samples in the test set, which consists of 24 hours of data. Their performance is summarized in Table~\ref{tab:prediction_summary} in terms of~\ac{MAE} (in minute), $\textrm{R}^2$ score, and average computation time (in second, per flexibility envelope). AdaBoost performs the best according to all error metrics, whereas the linear model performs even worse than the benchmark model. In terms of computation time, AdaBoost is the second fastest \ac{ML} model after the linear model, which makes it the most appropriate candidate. In comparison, the original calculation using one CPU process is estimated to take $20.75\pm 16.05$ 
seconds per flexibility envelope generally. 
The reduction is achieved because \ac{ML}-based approaches do not involve an iterative solution process. 

\begin{figure}[t!]
    \centering
    \includegraphics[width=\columnwidth]{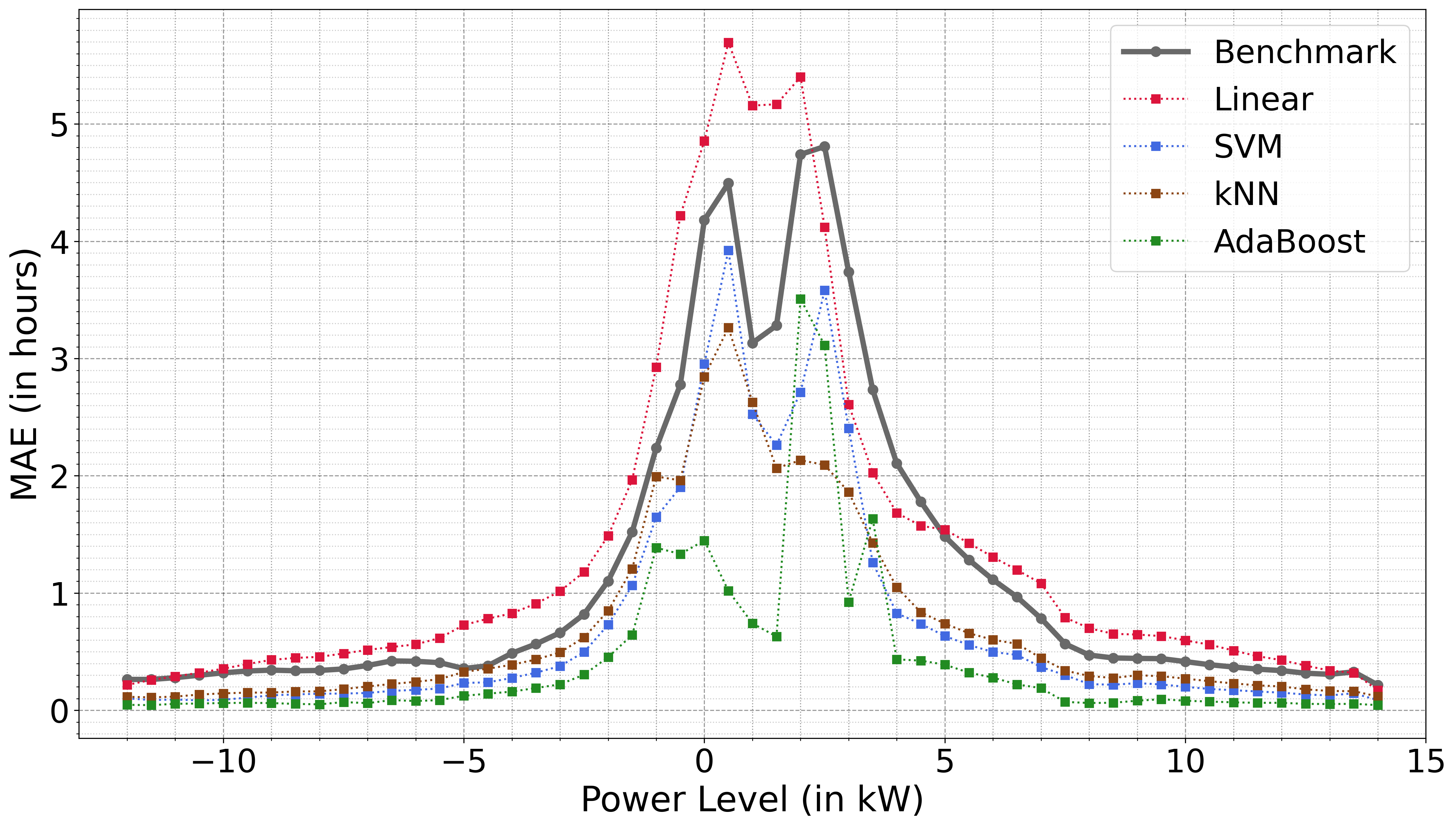}
    \caption{MAE of the flexibility prediction with respect to the power level.}
    \label{fig:prediction_MAE}
\end{figure}

\begin{figure}[t!]
    \centering
    \includegraphics[width=\columnwidth]{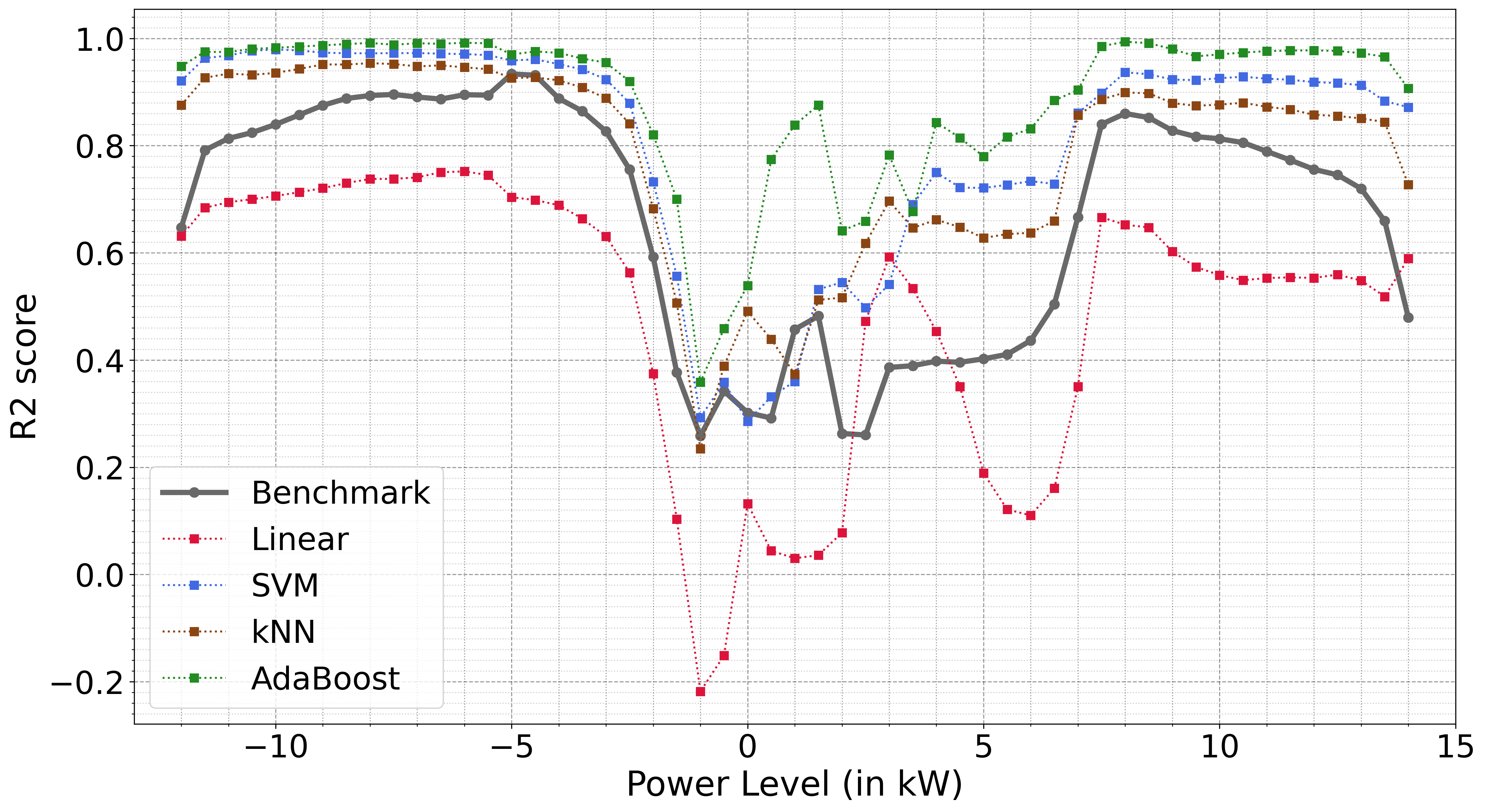}
    \caption{$\textrm{R}^2$ score of the flexibility prediction with respect to the power level.}
    \label{fig:prediction_R2}
\end{figure}

Moreover, Figs.~\ref{fig:prediction_MAE}, and~\ref{fig:prediction_R2} provide insights into the error distribution with respect to the power level. First of all, the ranking of performance observed in Table~\ref{tab:prediction_summary} among the various prediction models is respected for most of power levels. In terms of absolute error (i.e., MAE), power levels close to zero are naturally more impacted due to the generally higher magnitude.
Furthermore, the $\textrm{R}^2$ score indicates that the variance is generally harder to predict between the power levels of $-2$ kW and $8$ kW. Indeed, such power levels are a composition of small loads and PV output, characterized by high variance. In contrast, the flexibility outside $[-2,8]~\textrm{kW}$ is mainly attributed to the~\ac{EV}, which has more regular patterns.

\subsection{Approximation of Flexibility Envelope}

\begin{figure}[t!]
    \centering
    \includegraphics[width=\columnwidth]{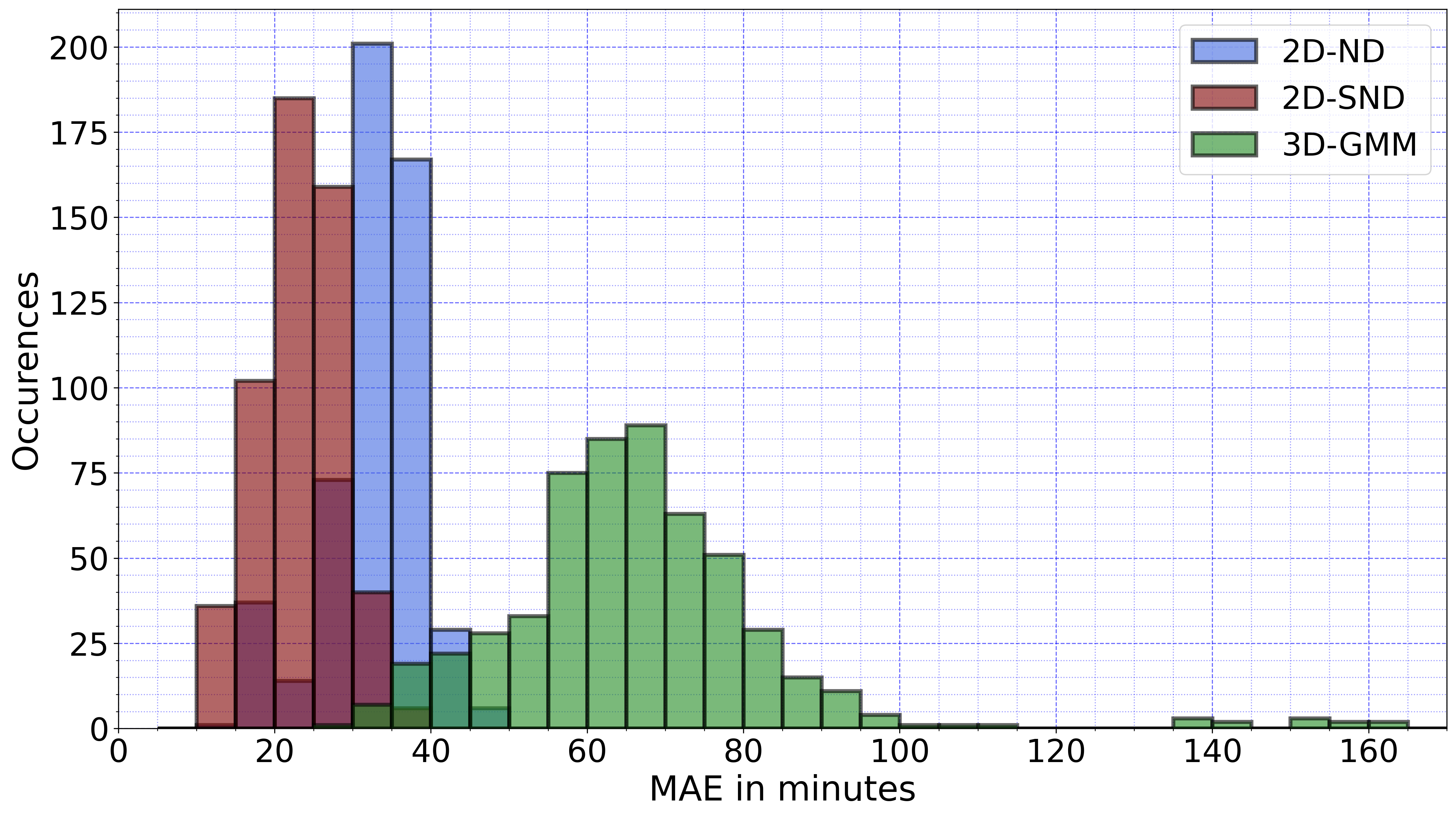}
    \caption{Distribution of the MAE for the flexibility envelope approximation.}
    \label{fig:approximation_MAE}
\end{figure}

\begin{figure}[t!]
    \centering
    \includegraphics[width=\columnwidth]{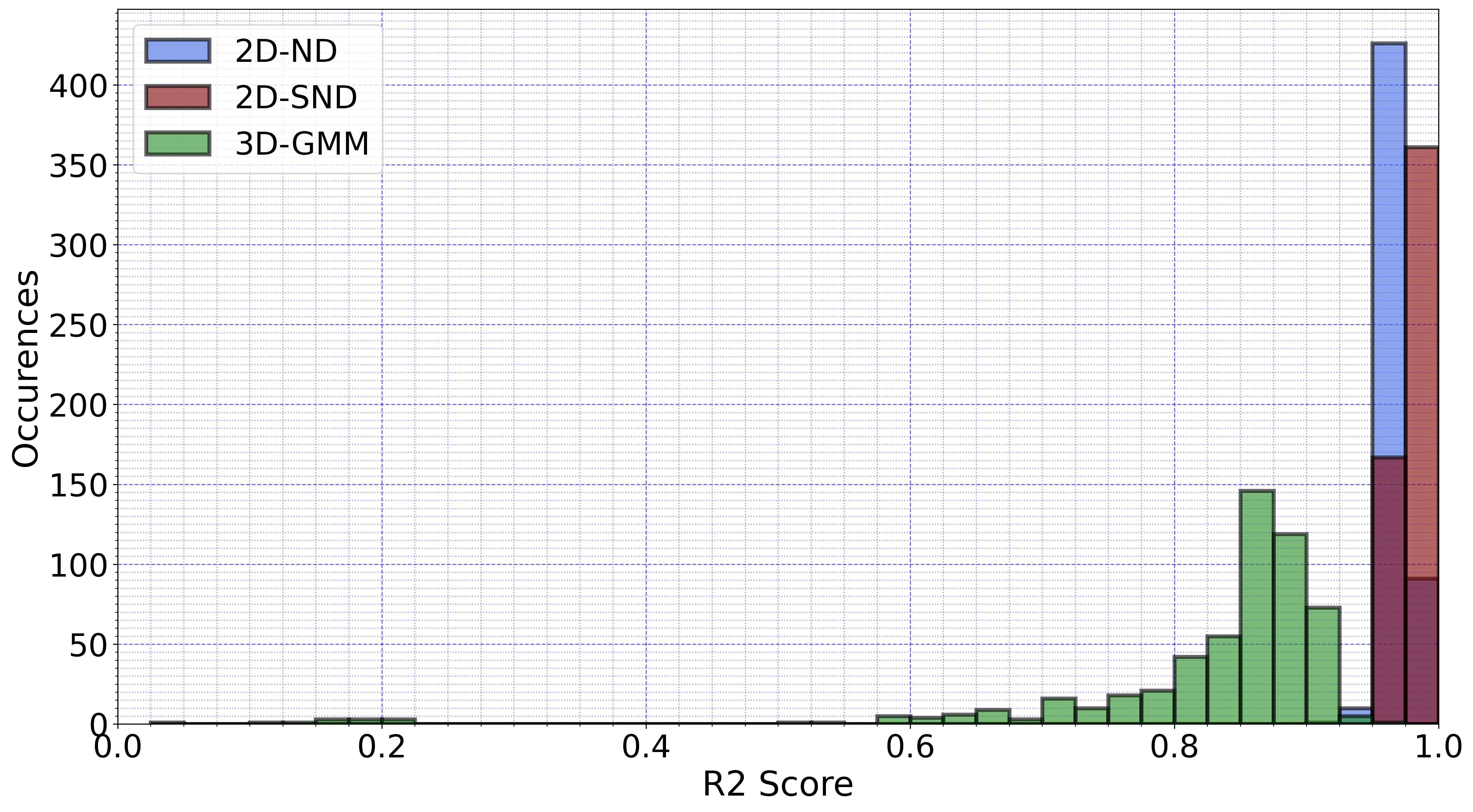}
    \caption{Distribution of the $\textrm{R}^2$ score for the flexibility envelope approximation.}
    \label{fig:approximation_R2}
\end{figure}

The three approximation models have been applied to all the flexibility envelopes in the dataset. Their performance is summarized in Table~\ref{tab:approximation_summary} in terms of number of parameters, average~\ac{MAE} (in minute), average $\textrm{R}^2$ score, and average computation time (in second) per flexibility envelope. In terms of accuracy, the 3-dimensional model is clearly outperformed by both 2-dimensional models. Indeed, each slice of envelope (i.e., corresponding to a given lead time) exhibits a distinctive Gaussian shape, but the specific parameters highly vary over the different lead times. Moreover, the skewed normal distribution performs slightly better than the standard version. We can observe from Figs.~\ref{fig:approximation_MAE} and~\ref{fig:approximation_R2} that the 3D model leads to a larger spread of the approximation errors. Note that the 2D models, and especially the~\ac{SND}, can almost perfectly account for the variance in the data. In contrast, the 3D-GMM outperforms both 2D approximators in terms of computational time (e.g., $8.7$ times faster than the 2D-ND) and required number of parameters. Compared to the original representation, the 2D-ND, 2D-SND, and 3D-GMM reduce the number of parameters by a factor of $8.83$, $6.63$, and $12.96$, respectively.
In fact, the best algorithm is a trade-off among approximation accuracy, computational and communication load.

\begin{table}[t!]
\caption{Performance summary of the approximation models.}
\label{tab:approximation_summary}
\centering
 \begin{tabular}{|c|c|c|c|c|} 
 \hline
 \begin{tabular}{@{}c@{}}Approx.\\model\end{tabular} &
 \begin{tabular}{@{}c@{}}Number of\\parameters\end{tabular} &
 \begin{tabular}{@{}c@{}}MAE\\(min)\end{tabular} &
 $\textrm{R}^2$ score &
 \begin{tabular}{@{}c@{}}Comput.\\time (sec)\end{tabular}\\ [0.5ex] 
 \hline
 2D-ND & 132 & 32 & 0.97 & 3.68\\
 2D-SND & 176 & \textbf{23} & \textbf{0.98} & 5.50\\ 
 3D-GMM & \textbf{90} & 66 & 0.83 & \textbf{0.63}\\ 
 \hline
\end{tabular}
\end{table}

\subsection{Overall Comparison}

\begin{figure*}
    \centering
    \includegraphics[width=\textwidth]{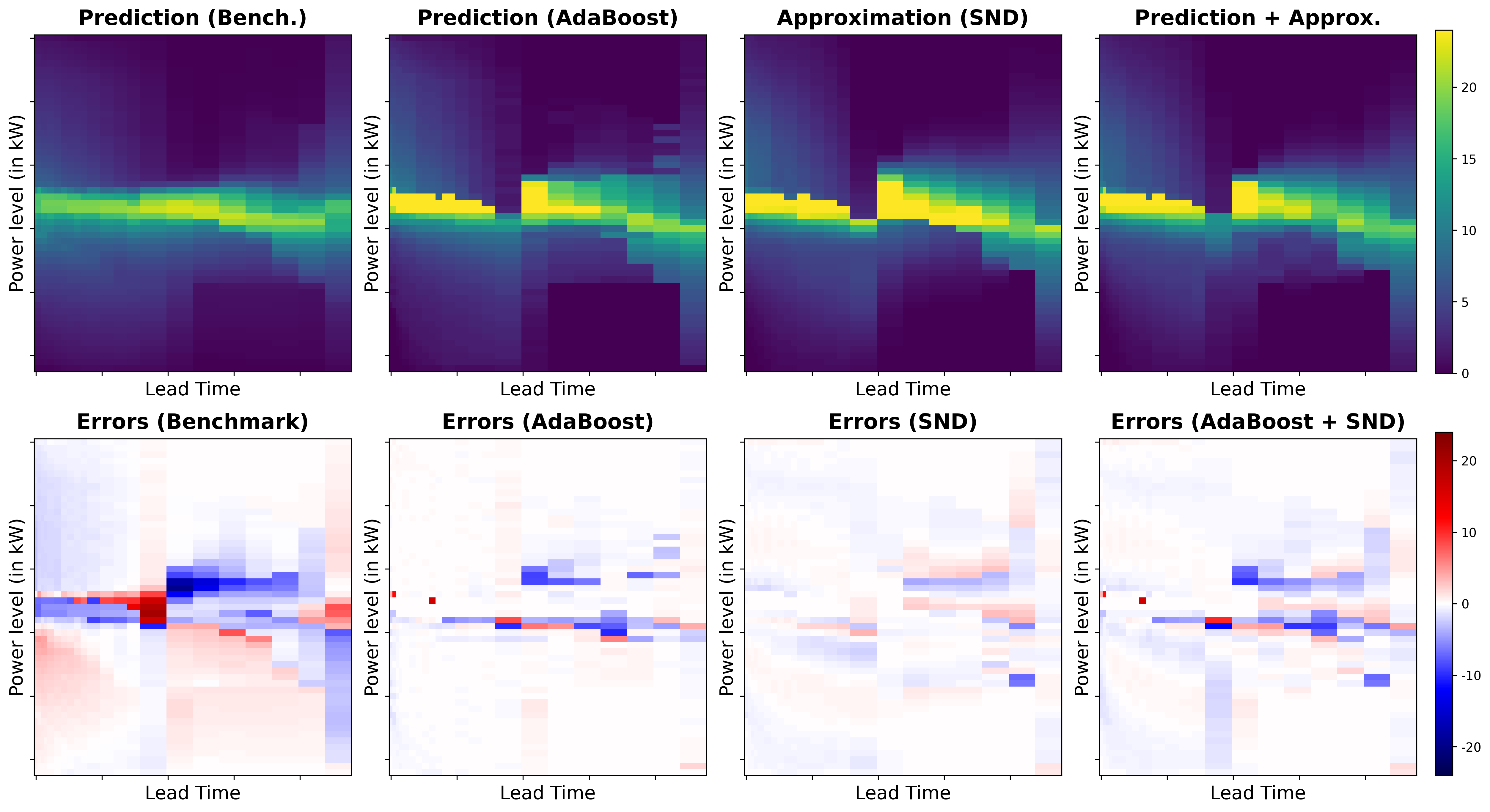}
    \caption{Example of the prediction and approximation of the flexibility envelope displayed in Fig.~\ref{fig:envelope}.}
    \label{fig:example}
\end{figure*}

Fig.~\ref{fig:example} shows the outcome of the benchmark and the best performing prediction (i.e., AdaBoost) and approximation (i.e., 2D skewed normal distribution) models for the example flexibility envelope from Fig.~\ref{fig:envelope}. The upper subplots display the predicted and approximated flexibility envelopes, whereas the lower subplots show the difference between the respective upper subplots and the original envelope. While the benchmark prediction model cannot account for the variation among different lead times, both best performing models provide an envelope visually close to the original envelope.
In addition, the best performing prediction and approximation models are integrated into a single pipeline, which is displayed in the right-hand side subplots. Formally, the 2D-SND approximates the envelope predicted by the AdaBoost algorithm. 
This corresponds to the desired process to reduce both the computational and communication load. 

Table~\ref{tab:pipeline_summary} summarizes the performance of the different models. As expected, the benchmark prediction model performs the worst. In absolute values, the entire envelope can be better predicted than approximated. However, the approximation still leads to higher $\textrm{R}^2$ score. Additionally, although the combination performs slightly worse than both models separately, the respective errors do not directly add up. This shows that these models can be used to speed up the quantification process and lower the communication load without much loss of accuracy.

Finally, a hybrid prediction model has also been considered. Precisely, instead of predicting all power levels based on the AdaBoost algorithm, model is selected in the training phase and the best performing prediction model per power level is applied on the test set (see Fig.~\ref{fig:prediction_MAE}). Consequently, AdaBoost is still used in most of the cases but is replaced by the~\ac{SVM} and~\ac{KNN} algorithms for some power levels close to zero. This hybrid approach achieves an MAE of 31 minutes (i.e., $-4$ minutes, compared to the accuracy displayed in Table~\ref{tab:pipeline_summary}), and a $\textrm{R}^2$ score of $0.95$ (i.e., $+0.01$) for the combination.

\begin{table}
\caption{Performance summary of the prediction and approximation models on the flexibility envelope displayed in Fig.~\ref{fig:envelope}}
\label{tab:pipeline_summary}
\centering
 \begin{tabular}{|c|c|c|} 
 \hline
 Model & MAE (min) & $\textrm{R}^2$ score \\ [0.5ex] 
 \hline
 Prediction (Benchmark) & 93 & 0.79 \\
 Prediction (AdaBoost) & 21 & 0.95 \\ 
 Approximation (SND) & 25 & 0.99 \\ 
 Combin. (AdaBoost + SND) & 35 & 0.94 \\ 
 \hline
\end{tabular}
\end{table}


\subsection{Computational Aspects}

The original quantification was performed on a server with an Intel Core i7-8565U CPU and 32~GB of RAM. The optimization problem was solved with CVXPY and MOSEK.
For the \ac{ML}-based quantification, another server with an Intel Core i7-8565U CPU and 16~GB of RAM is used  and the results are obtained using the scikit-learn library.

Note that the time spent on offline training is not accounted for since it does not influence the real-time flexibility quantification process. 
However, training is again required if a new flexible component is added or if there are substantial changes in the flexibility quantification model or user behavior.

\section{Conclusion and outlook}\label{sec:conclusion}
This paper presents a data-driven combination of prediction and approximation to carry out energy flexibility quantification and communication efficiently, which will facilitate real-time and large-scale flexibility utilization. The results show that the proposed pipeline significantly reduces the computation and communication burdens of the online stage, while obtaining satisfactory approximation. Nonetheless, the proposed approach still relies on the existing approach to collect data for training. In the future, we will investigate sample efficiency and the possibility of removing the current ad-hoc data collection process to enhance scalability.

\section*{Acknowledgement}
The work of H. Cai and P. Heer was supported by the Swiss Federal Office of Energy through the Sustainable Demand-Side Management for the Operation of Buildings (S-DSM) project under the contract number SI/502165-01. The work of T. Zufferey and G. Hug is part of the activities of the Renewable Management and Real-Time Control Platform (ReMaP).

\bibliographystyle{IEEEtran}
\bibliography{reference}

\end{document}